%% file: zz_main.tex
\documentclass[12pt]{article}

\usepackage{sbc-template}

\usepackage{graphicx,url}
\usepackage{amsmath, amssymb}
\usepackage{float}
\usepackage[utf8]{inputenc}  
\usepackage{indentfirst}
\usepackage{natbib}
\usepackage{multicol}
\usepackage[inline]{enumitem}
\usepackage{booktabs}
\usepackage{lscape}
\usepackage{subfig}

\usepackage{tikz}
\usetikzlibrary{decorations.pathmorphing}

\usepackage[textsize=scriptsize]{todonotes}
     
\title{%Multi-step ahead
MegazordNet: combining statistical and machine learning standpoints for time series forecasting
}

\author{Angelo Garangau Menezes\inst{1*}, Saulo Martiello Mastelini\inst{1}\thanks{Authors have contributed equally.}}

\address{
  Instituto de Ciências Matemáticas e de Computação -- Universidade de São Paulo\\
  Av. Trabalhador São Carlense, 400 -- 13566-590, São Carlos -- SP, Brasil.
  \email{angelomenezes.eng@gmail.com, saulomastelini@gmail.com}
}

\begin{document} 

\maketitle

\begin{abstract}

Forecasting financial time series is considered to be a difficult task due to the chaotic feature of the series. Statistical approaches have shown solid results in some specific problems such as predicting market direction and single-price of stocks; however, with the recent advances in deep learning and big data techniques, new promising options have arises to tackle financial time series forecasting. Moreover, recent literature has shown that employing a combination of statistics and machine learning may improve accuracy in the forecasts in comparison to single solutions. Taking into consideration the mentioned aspects, in this work, we proposed the MegazordNet, a framework that explores statistical features within a financial series combined with a structured deep learning model for time series forecasting. We evaluated our approach predicting the closing price of stocks in the S\&P 500 using different metrics, and we were able to beat single statistical and machine learning methods.

\end{abstract}

\input{sections/01_introduction.tex}
\input{sections/02_related_works.tex}
\input{sections/03_proposed_approach.tex}
\input{sections/04_materials_and_methods.tex}
\input{sections/05_results_and_discussion.tex}
\input{sections/06_final_considerations.tex}

%\begin{figure}[!htb]
	%\begin{center}
	%	\includegraphics[width=\textwidth]{stock.pdf}
	%\end{center}
    %\caption{\label{fig:stock} Ilustração das ações da empresa `3M' nos últimos 5 anos.}
%\end{figure}

\bibliographystyle{apalike}
\bibliography{sbc-template}

\end{document}

%% file: sections/01_introduction.tex
\section{Introduction}
%\todo[inline]{Acho que vale a pena dar uma ênfase na ideia que utilizamos simples ideias de um statistical standpoint combinadas com state-of-the-art ML e conseguimos obter resultados interessantes. Aí deixamos como trabalhos futuros analisar outras configurações, i.e., ajustes nos hyperparâmetros do nosso framework}

The idea of forecasting the future has gained the attention of researchers through the time~\citep{brockwell2002introduction}. The impact of this achievement is manifold: in some areas, this implies in obtaining useful knowledge about the environment or better comprehending behaviors, such as one of the populations. In the context of financial markets, the advantage of being able to forecast tendencies and movements is clear and specific: it means money. In fact, since the conception of stock markets, researchers are trying to find ways of describing and modeling the complex phenomena involved in this kind of time series (TS)~\citep{tsay2000time, brockwell2002introduction}.

As the technology evolved along with the sophistication in mathematical and statistical modeling, the approaches for financial time series forecasting (FTSF) also became more accurate. Several research areas tackle this prediction problem using different theoretical basis and strategies. They include, for instance, statistical modeling~\citep{brockwell2002introduction,makridakis2018m4,parmezan2019evaluation}, Machine Learning (ML) approaches~\citep{makridakis2018statistical,parmezan2019evaluation}, Evolutionary Programming~\citep{aguilar2015genetic}, among others. The increased popularity of deep learning (DL) pushed forward, even more, the possibilities for time series forecasting (TSF)~\citep{nelson2017stock,parmezan2019evaluation}.

While many general-purpose and successful TSF methods have been proposed in the past years, the specific branch of financial time series brings multiple additional difficulties. Firstly, this type of TS is strongly dependent on external agents, such as the political situation, the public opinion about a company, natural disasters, among other aspects~\citep{hu2018predicting,parmezan2019evaluation}. Secondly, financial TS commonly do not present stationarity, an important aspect which many statistical methods assume as a prerequisite for application~\citep{parmezan2019evaluation}. The mentioned aspects lead to traditional TS techniques failing to capture stock movements, and thus, not offering a reliable tool for investment planning. %This is even worse when considering multi-step-ahead prediction since the accuracy deteriorates as the prediction horizon increases~\citep{parmezan2019evaluation}.

A lengthy debate concerning which type of prediction technique is most suitable for FTSF has been held by researchers with ML and statistics backgrounds~\citep{chen2015lstm,nelson2017stock,bao2017deep,hu2018predicting, makridakis2018statistical}. Unfortunately, most of these works are limited by only considering a few techniques, neglecting others, or even not exploring the full potential of the compared methods.  On the other hand, recent literature points out that the usage of hybrid approaches tends to lead to the best results~\citep{makridakis2018m4}. Moreover, the investor may be interested in other aspects than just decreasing the error in the obtained forecasts. For example, one may wish to only describe up and down trends in the long term, instead of predicting the next day accurately.

In this sense, combining the best tools of different research areas seems to be an appealing way to tackle FTSF. In this work, we investigate the usage of state-of-the-art TSF techniques for financial data. We consider both statistical and ML approaches to address the one-step-ahead forecasting task and propose a simple framework for relevant financial data feature extraction and regression called MegazordNet. Specifically, we investigate and apply our methods to $148$ randomly chosen TS from the S\&P 500 index; nevertheless, our ideas can be extended for other financial scenarios and markets.

This work is divided as follows: Section \ref{related-work} gives an overview of different ML and statistical approaches to the one-step-ahead TS forecasting task; Section \ref{proposed-approach} presents the inner-workings of the developed approach; Section \ref{materials-and-methods} describes the methodology used along with the resources and the experimentation setup; Section \ref{results} discusses our results and findings followed by the final considerations and future work indicated in Section \ref{final-considerations}.

%% file: sections/02_related_works.tex
\section{Related Work}\label{related-work}

Even with the predominance of statistical modeling along the years, ML has currently been applied vastly to the context of FTSF as can be seen by different surveys on the topic~\citep{chong2017deep, chatzis2018forecasting, patel2015predicting}. Statistical approaches are often compared with DL and more traditional regression methods, but their performance may be highly dependent on the solved problem since they all depend on the quality and amount of data available for the task~\citep{chen2014big}.

\cite{parmezan2019evaluation} evaluated different statistical and ML algorithms for TSF using 40 synthetic and 55 real datasets. According to the obtained results for various metrics such as MSE, Theil's U coefficient, and POCID, statistical approaches were not able to outperform ML-based techniques with a statistical difference. One of the authors' contributions was the organization of a repository that contains all datasets used on their analysis in order to facilitate study replication and evaluation for other modeling techniques.

\cite{makridakis2018m4} presented the results of the M4 competition whose goals were to research new ways for improving TSF accuracy, and how such learning can be applied to advance the theory and practice of forecasting. The competition presented statistical, ML and ``hybrid'' approaches for modeling complex TS data from different fields such as tourism, trade, and wage. The paper reinforces the idea that a single technique may not be suited to all problems, but the combination of some usually are able to bring satisfactory results.

Regarding the specific problem of predicting the direction of the US stock market, the work of \cite{hu2018predicting} evaluated different optimization techniques for determining the optimal set of parameters for an artificial neural network (ANN) created to model the trends of the US stock market. They used data from the S\&P500 and DJIA Indices along with Google Trends data to model the TS. Their results showed the impact of exploring not only TS values, but also different external sources based on the sentiment of the general public and investors such as Google Trends for financial forecasting.

%\cite{nguyen2015topic} explored different sources of data to create a model that could predict the price movement of 5 US stocks. They were able to improve a naive model that used only historical data over 6\% in accuracy using sentiment analysis on data from message boards of the stocks from Yahoo Finance Message Board where people discuss news, comments, and statistics related to these companies. This approach supports the hypothesis that the price of a stock may be not only dependent on past values but also on the general public sentiment about the company.

\cite{bai2018empirical} compared the results of networks with simple convolutional architectures against canonical recurrent neural networks (RNN) to check which architecture would have better performance. The latter type of ANNs include, for instance, the Long-Short Term Memory (LSTM) networks. The performed analysis comprehended a diverse range of tasks, including TSF and sequence modeling. Their findings support the idea that networks with convolutional architectures present longer sufficient memory and may also be used as a benchmark for such types of problems. 

\cite{lin2017hybrid} proposed a pipeline where a convolutional neural network (CNN) extracts salient features from local raw data of TS while a LSTM network models existing long-range dependency within the trend of the historical data. They used a feature fusion dense layer to learn a joint representation able to predict the trend over the TS. The authors were able to achieve better results than vanilla LSTM, Conv-LSTM, Support Vector Regression, and other simpler ML models in three different TS datasets.

Most of the existing solutions focus only on a specific branch of forecasting research, such as ML or statistical modeling. Hence, the solutions employ a single point-of-view over the TS modeling. This strategy can be sub-optimal in tackling a complex task such as TSF. We hypothesize that combining the description power of the statistical methods and the learning capabilities of DL can lead to a better predictive performance in FTSF.

%% file: sections/03_proposed_approach.tex
\section{Proposed approach}\label{proposed-approach}

\begin{figure}[!htb]
    \centering
    \includegraphics[width=0.48\textwidth]{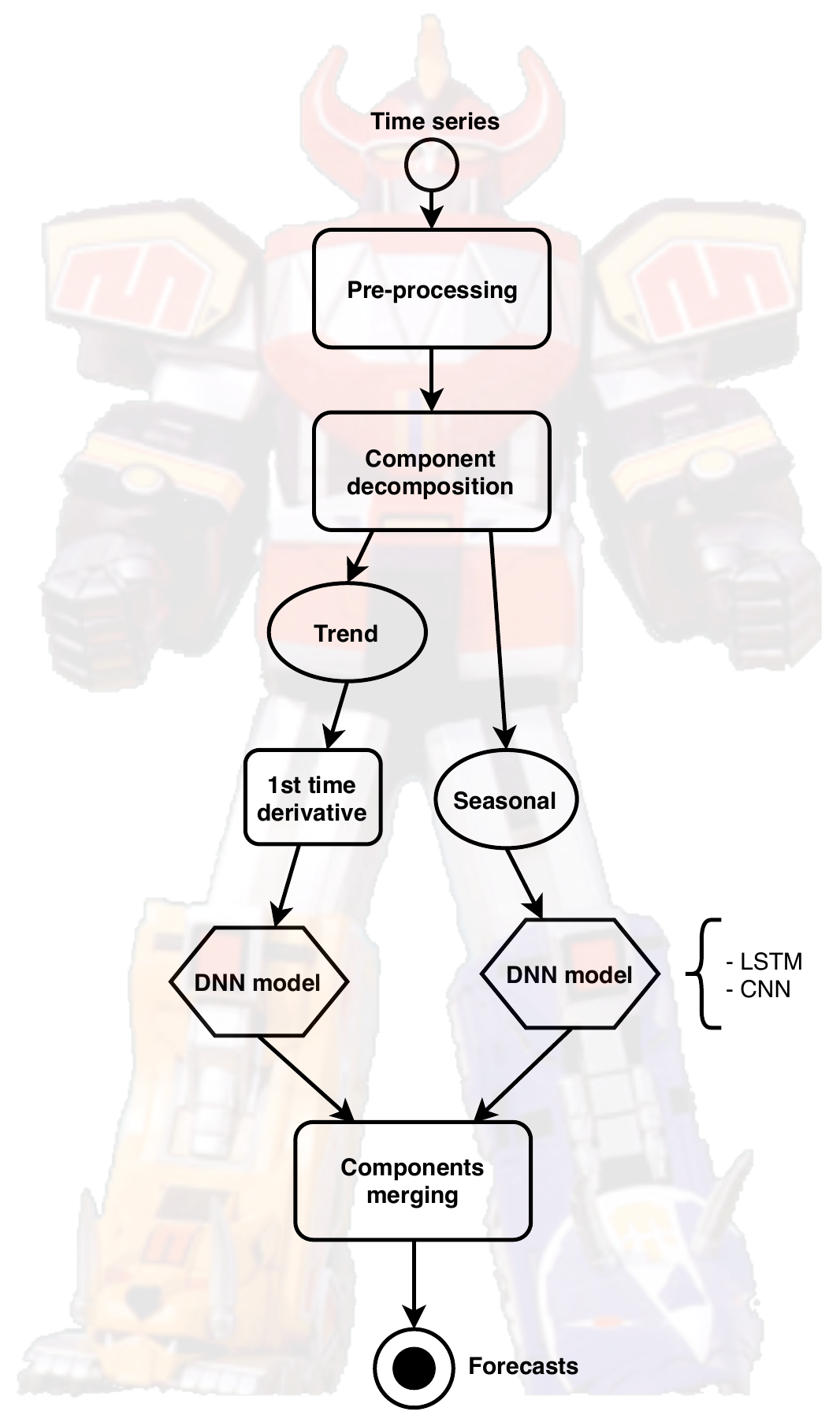}
    \caption{Overview of MegazordNet}
    \label{fig:overview}
\end{figure}

In this section, we present our proposal, called MegazordNet, for TSF. In this work, we focus on FTSF, but our ideas can be easily extended to other TS domains. Figure~\ref{fig:overview} presents an overview of our proposed approach for TSF. Firstly, for evaluation, we divide the TS of interest in training and testing portions. The former portion is employed for inducing the prediction models while the latter is responsible for evaluating the predictors in scenarios they have never seen before, i.e., when making predictions for some steps in the future. In this context, when partitioning the TS, the time order must be kept. As we want to make predictions for the closing price of the next day, we treat this as an univariate problem using only this attribute for the TS modeling. This, again, can be easily extended to the multivariate cases.

\subsection{Pre-processing and time series component decomposition}

After obtaining the training partition, the input data is subjected to a pre-processing step. At this point, we aim at improving the data representation that the algorithms will receive at the next step. In our proposal we just removed missing entries from the TS. Following, the inputs are decomposed into Trend and Seasonal components~\citep{parmezan2019evaluation}. Since financial TS represent complex data patterns, often influenced by external factors, decomposing the original series into different components ought to make the data representation easier to be modelled by the prediction algorithms~\citep{hu2018predicting}. Component decomposition in TS analysis is a common practice performed from a statistical point-of-view. Nevertheless, most of the machine learning approaches for tackling TSF seem to ignore this kind of mechanism. \citep{chen2007flexible, wen2019stock, chong2017deep} 

\begin{figure}[!htb]
    \centering
    \includegraphics[width=0.8\textwidth]{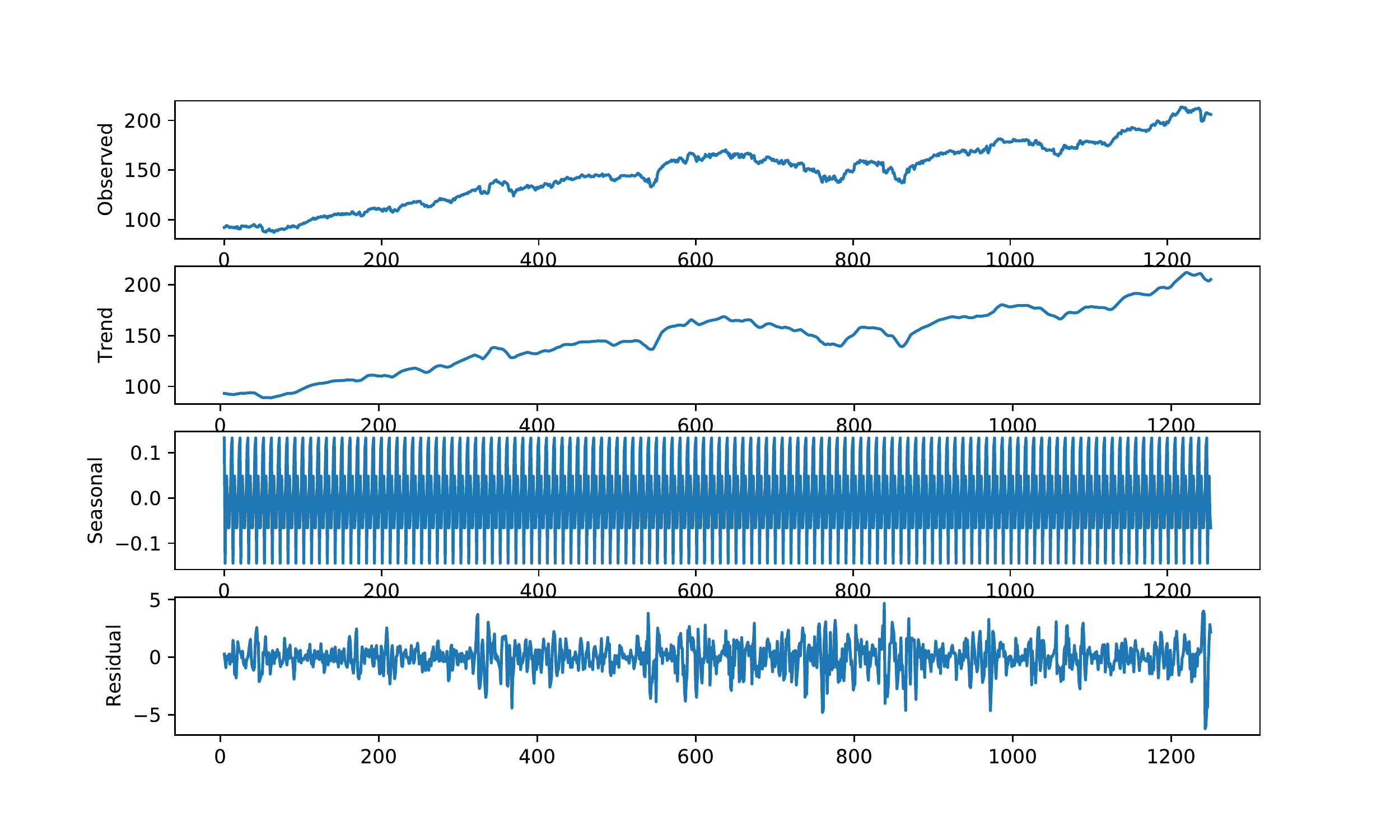}
    \caption{``3M'' time series components}
    \label{fig:decomp}
\end{figure}

We used a simple moving average approach for trend, seasonal, and residual components extraction. An empirically determined window size of $10$ days was employed to this end. An example of the applied operation can be seen in Figure \ref{fig:decomp}. After decomposition, we modeled the trend and the seasonal components separately in order to learn the best model fit for each of them, and get individual forecasts. Moreover, given the non-stationary characteristics of financial TS, we applied the first time derivative for the trend component. In this way, we allow our trend model to learn only the variations from a time observation to the other. Thus, for the final trend forecasting, MegazordNet adds the outcome of its trained trend-variation model to the previous trend observation.

We chose not to model the residual component since financial stocks present many small chaotic fluctuations which could disturb the final results of the proposed approach~\citep{chen2007flexible}. With the forecast of trend and seasonal components, we are able to apply an additive model and get the prediction for the following time step as the sum of the separate components forecasts. 

\subsection{Component forecasting}

As we wanted to explore the state-of-the-art for TSF, we employed in this work two-variants of neural networks that have been commonly used to sequence-to-sequence modeling: CNNs and LSTMs.

\subsubsection{CNN}

A convolutional neural network is a biologically-inspired type of deep neural network that maps the output to a local region of the inputs through the multiplication of a sliding weight matrix (weight filters). These filters update their values, usually via an optimization technique that minimizes some loss based on the difference between predicted and expected outputs. Intuitively, the idea of applying CNNs to time series forecasting would be to learn filters that represent specific repeating patterns in the series and use these to forecast the future values~\citep{borovykh2017conditional}.

Even though only in the last few years CNNs have proved to be an essential tool for TS forecasting, researchers more than 20 years ago had already alerted to their effectiveness in these kind of problems~\citep{lecun1995convolutional}. This architecture is particularly suitable for modeling long TS because, depending on its hyperparameters, it may have a greater receptive field than a typical recurrent network, which has been the ``go to'' architecture for TSF~\citep{shen2019novel}.

\begin{figure}[!htb]
    \centering
    \includegraphics[width=.5\textwidth]{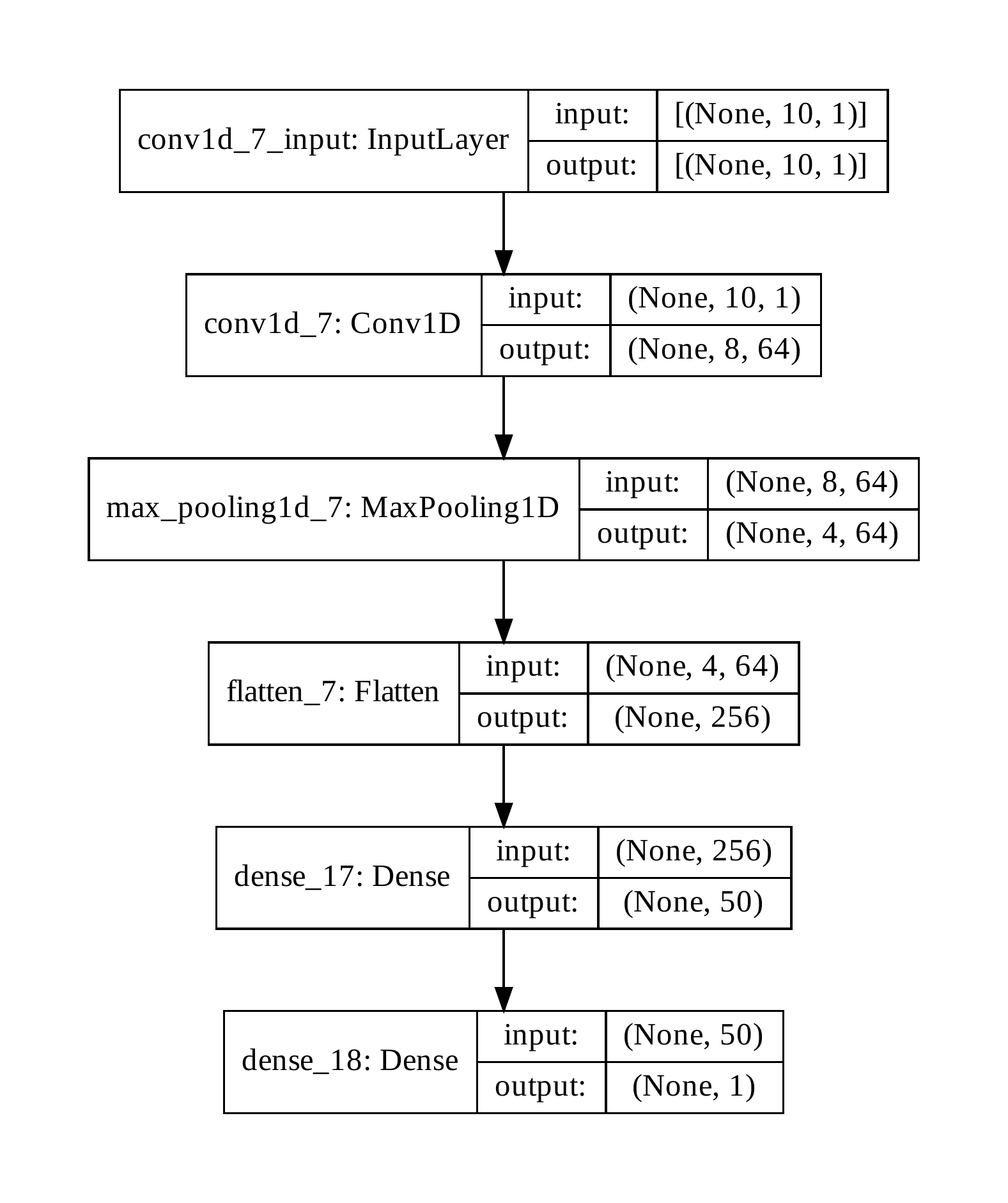}
    \caption{CNN architecture used}
    \label{fig:cnn}
\end{figure}

As our base CNN model, we used a convolutional layer with 64 filters and kernel size 3, followed by a max-pooling layer, a flatten layer, and two dense layers. A visualization of the employed architecture is shown in Figure~\ref{fig:cnn}. All the used CNNs were trained for 100 epochs using the Adam optimizer with a learning rate of $0.001$.

\subsubsection{LSTM}

Recurrent neural networks have been applied massively to TSF since they own a specific structure that deals with sequences and ordered data. This type of ANN gives as an input to each neuron not only the next data point but also the value of the previous state. Such strategy creates a notion of memory and recursive feedback to the learning model. Since some financial TS are well-known for their autoregressive feature, i.e., the next time step has a reasonable probability of being correlated with its last state, RNNs present themselves as a reasonable modeling solution~\citep{chen2008high}.

One problem that arises from the training process of a RNN is that the gradient of some of the weights often starts to become too small or too large when the network has a large receptive field (``memory''). This causes the network not to update its weights properly and is called the vanishing gradients problem~\citep{bengio1994learning}. A type of network architecture that solves this problem is the LSTM, wherein a typical implementation, the hidden layer is replaced by a complex block of computing units composed by gates that trap the error in the block, forming a so-called ``error carrousel''~\citep{gamboa2017deep}.

\begin{figure}[!htb]
    \centering
    \includegraphics[width=.5\textwidth]{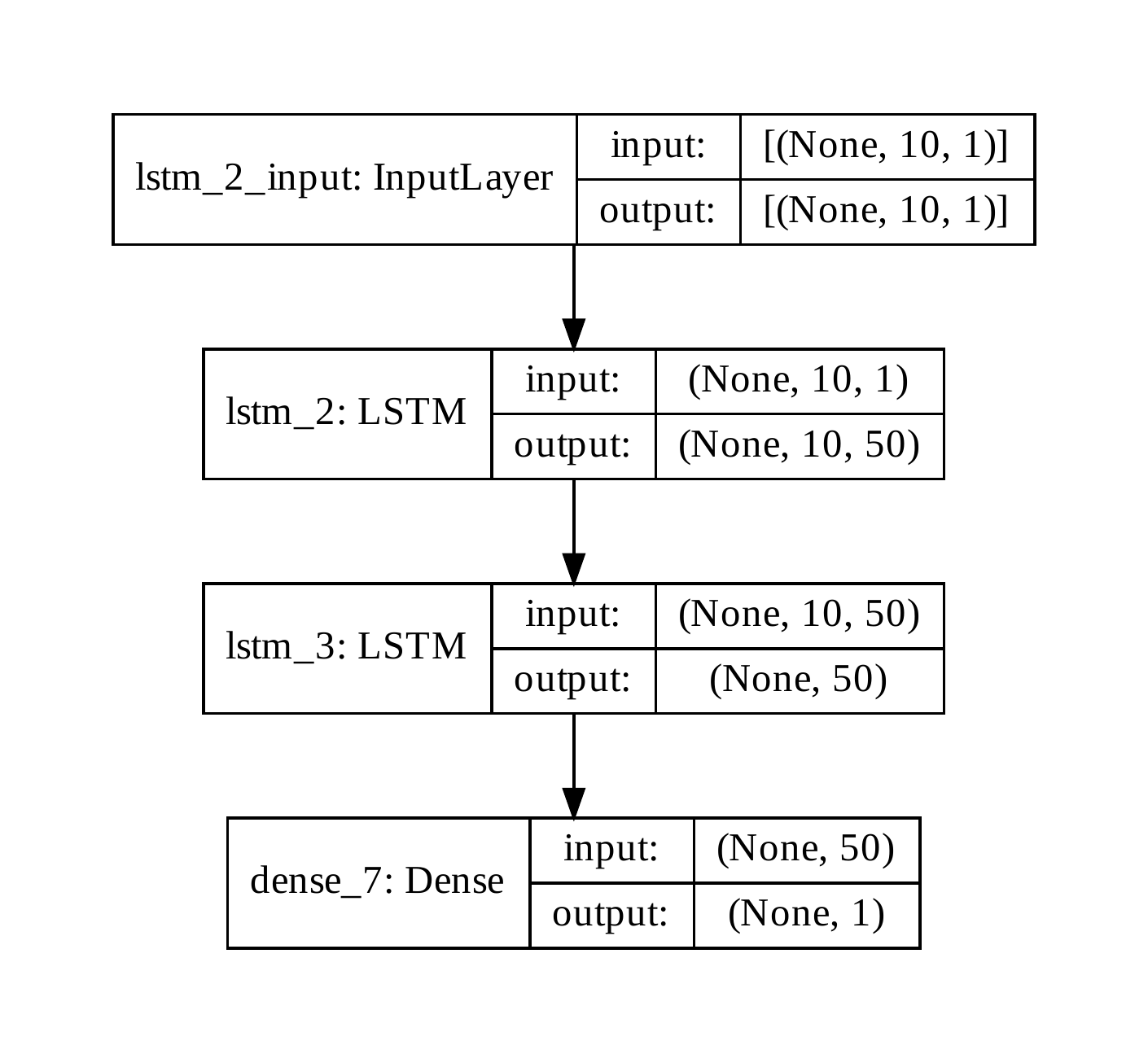}
    \caption{RNN architecture used}
    \label{fig:rnn}
\end{figure}

Using the LSTM architecture as our recurrent base model, we stacked two LSTM layers with $50$ units and a dense layer. A sample of the architecture can be seen in Figure~\ref{fig:rnn}. The networks were trained for $100$ epochs using the Adam optimizer with a learning rate of $0.001$.

%% file: sections/04_materials_and_methods.tex
\section{Materials and Method}\label{materials-and-methods}

This section describes the resources and methodology we have used to deal with the TSF task.

\subsection{Data description}
The S\&P 500 dataset presents a period of five years in economic transactions for the S\&P 500 index. This index comprehends the $503$ most economically prominent companies in US. For each company, approximately $1258$ daily observations were recorded. These records describe the variations in the index's actives shares in the stock market. In this sense, the data samples can be seen as TS, for mathematical and computational modelling.

Altogether, $606,800$ samples compose the mentioned dataset. In fact, different companies do not have the same number of observations, given the vast time range considered, i.e., some companies may left or enter the top $500$ ranking, or even go bankrupt. Besides, for some records, there are missing information in their description features. By removing these incomplete samples, the total number of data records decreases to $601,011$. Table~\ref{tab:dataset_description} presents the features contained in the mentioned dataset.

\begin{table}[!htb]
    \centering
    \caption{S\&P 500 properties description}
    \begin{tabular}{cl}
        \hline
        \textbf{Type} & \textbf{Description} \\ \hline
        \textit{Date} & Observation date \\
        \textit{Open} & Open price for the active \\
        \textit{High} & Maximum price reached by the active \\
        \textit{Low} & Minimum price reached by the active\\
        \textit{Close} & Closing price for the considered active \\
        \textit{Volume} & Number of transactions for the active \\
        \textit{Name} & Company acronym \\
        \hline
    \end{tabular}
    \label{tab:dataset_description}
\end{table}

We considered a total of $148$ companies from the S\&P 500 group in our analysis. Those TS were selected randomly to provide a general standpoint for financial forecasting analysis. In the future we intend to extend our experimental analysis to all the companies in the dealt index.

\subsection{Experimental setup}

Aiming at evaluating the performance of the different approaches for TSF, we employed the \textit{holdout} data partition scheme, as suggested in recent literature~\citep{makridakis2018statistical, parmezan2019evaluation}. For this end, we employed an 80/20 strategy, i.e., $80\%$ of the TS were employed for inducing the prediction models, whereas the remaining $20\%$ were reserved for evaluating the predictors.

It is worth mentioning that the other methods included in our comparison use rolling windows to update their parameters, i.e., they employ an online learning mechanism. Nevertheless, all the compared algorithms were evaluated using the same portions of the TS. We intend to evaluate the possibility of applying online learning for the MegazordNets in the future.

\subsubsection{Algorithms compared against our proposal}

Aiming at comprising a wide range of algorithmic solutions to compete against our proposal, we selected well-known statistical and ML-based algorithms for TSF. In the group of statistical based algorithms, we selected three autoregressive models, one exponential smoothing solution, and one straightforward moving average predictor. We also included a recent ML solution, which is based on neighborhood information~\cite{parmezan2019evaluation}.

Table~\ref{tab:contender_algorithms} summarizes the algorithms that were compared against MegazordNet in this study, along with their settings. We fixed the hyperparameter settings for both MegazordNet and the compared methods regardless of the TS considered. We found empirically that those settings led to satisfactory results in most of the cases. In this sense, we aimed at performing a fair comparison of the algorithms regarding their robustness to several evaluation scenarios. 

In the table, the tuple following the ARIMA variants in the form $(p,d,q)$ refers to the traditional way of representing this TS forecasting algorithm. In this representation, $p$ represents the order (number of time lags) of the regressive model, $q$ represents the degree of differentiation, and $q$ the order of the moving average model. Moreover, $\alpha$ represents the decaying factor for SES, $w$ the window of time intervals considered in MA and k-NN-TSP, and $k$ the number of neighbors employed by k-NN-TSP.

\begin{table}[!htb]
    \centering
    \caption{Algorithmic solutions evaluated against our proposal.}
    \label{tab:contender_algorithms}
    \resizebox{\textwidth}{!}{
    \begin{tabular}{lp{0.5\textwidth}lc}
        \toprule
        \textbf{Acronym} & \textbf{Name} & \textbf{Settings} & \textbf{References} \\
        \midrule
        ARIMA & Autoregressive Integrated Moving Average & ARIMA$(1,1,0)$ & \cite{bhowmick2019performance} \\ \hline
        AR & Autoregressive model & ARIMA$(1,0,0)$ & \cite{kandananond2012comparison} \\ \hline
        RW & Random Walk & ARIMA$(0,1,0)$ & \cite{kandananond2012comparison} \\ \hline
        SES & Simple Exponential Smoothing & $\alpha=0.95$ & - \\ \hline
        MA & Moving Average & $w=10$ & - \\ \hline
        k-NN-TSP & k-Nearest Neighbours for Time Series Prediction & $w=5$, $k=3$ & \cite{parmezan2019evaluation} \\
        \bottomrule
    \end{tabular}}
\end{table}

\subsubsection{Evaluated MegazordNet variants}

As previously discussed, in this preliminary study, we considered two types of neural networks for TSF: LSTM and CNN. Given that MegazordNet builds different predictors for both trend and seasonal components, four different combinations of those neural networks could be assembled. Additionally, we also evaluated the possibility of only using the trend component for making the forecasts. Thus, two additional MegazordNet variants were considered. We evaluated all of those configurations when comparing our proposal against other TSF algorithms. The acronyms for the MegazordNet variants are presented in Table~\ref{tab:megazord_variants}, along with their meaning.

\begin{table}[!htb]
    \centering
    \caption{Evaluated MegazordNet variants and their acronyms}
    \begin{tabular}{ccc}
        \hline
        \textbf{Acronym} & \textbf{Trend Component} & \textbf{Seasonal Component} \\
        \hline
        $\text{MegazordNet}_{\text{L,L}}$ & LSTM & LSTM \\
        $\text{MegazordNet}_{\text{L,C}}$ & LSTM & CNN \\
        $\text{MegazordNet}_{\text{L,0}}$ & LSTM & - \\
        $\text{MegazordNet}_{\text{C,L}}$ & CNN & LSTM \\
        $\text{MegazordNet}_{\text{C,C}}$ & CNN & CNN \\
        $\text{MegazordNet}_{\text{C,0}}$ & CNN & - \\
        \hline
    \end{tabular}
    \label{tab:megazord_variants}
\end{table}

\subsection{Evaluation metrics}\label{sec:metrics}

Regarding the performance measurements, three well-known metrics were employed, as suggested by \cite{parmezan2019evaluation}, which were namely: Mean Square Error (MSE), Theil's U (TU) coefficient, and the hit rate Prediction of Change in Direction (POCID). In the following definitions, $h$ represents the forecasting horizon (the number of examples which are used for evaluating the prediction models), $z$ and $\hat{z}$ represent, respectively, the TS and its predicted values.

MSE measures the squared deviations of the predicted responses from the expected forecasts. In this sense, this metric smooths amounts lying within the range $[0,1]$ whereas accentuating differences greater than one. The MSE calculation is given by Equation~\ref{eq:mse}. 

\begin{equation}
    \text{\textit{MSE}} = \frac{1}{h} \sum_{t=1}^{h} (z_t - \hat{z}_t)^2
    \label{eq:mse}
\end{equation}

Sometimes, just observing the deviations from the expected responses is not enough to infer whether or not a model was accurate in its responses. For instance, when comparing two different TS, one from a company which plays an important role in the country's economy and another which is just beginning to operate in the stock market, the scale of their stocks tend to be very different. One option is to transform both TS into a common scale, e.g., to make them lie within the $[0,1]$ range. Notwithstanding, this option often difficulties the interpretation of the obtained results, considering that errors are not anymore computed in terms of monetary losses or gains. Another option is to compare the predictions of a model with the ones obtained from a baseline predictor, observing if gains were obtained. 

In this context, the TU metric compares the predictions of a model with a naive predictor which always outputs the immediately previous observation. Therefore, if $TU > 1$, the compared model was less accurate than the naive model. If $TU = 1$, the compared predictor behaved exactly as the naive model. Anytime $T < 1$, performance gains were obtained. TU's calculation is presented in Equation~\ref{eq:tu}.

\begin{equation}
    \text{\textit{TU}} = \frac{\sum_{t=1}^h (z_t - \hat{z}_t)^2}{\sum_{t=1}^h (z_t - z_{t-1})^2}
    \label{eq:tu}
\end{equation}

Lastly, we also can account for the amount of times a method was able to correctly predict the directions of change in the stock index, i.e., whether they would increase or decrease. For this end, we utilized the POCID metric which is given by Equation~\ref{eq:pocid}.

\begin{minipage}{.2\textwidth}
    \begin{equation*}
        \text{\textit{POCID}} = \frac{\sum_{t=1}^{h} D_t}{h} \times 100
    \end{equation*}
\end{minipage}
\begin{minipage}{.7\textwidth}
    \begin{equation}
        D_t = 
        \begin{cases}
             \text{1,} & \text{if} (\hat{z}_t - \hat{z}_{t-1})(z_t - z_{t-1}) > 0\\
             \text{0,} & \text{otherwise}
        \end{cases}
        \label{eq:pocid}
    \end{equation}
\end{minipage}

%Commonly, POCID is employed as a complementary metric in standard time series evaluation. However, in our context this metric is specially useful, given that we are interested in making multi-step ahead planning for stock investments.

%% file: sections/05_results_and_discussion.tex
\section{Results and Discussion}\label{results}

Considering that the companies present different ranges in their stocks, i.e., their worth can significantly differ, we mainly focused our discussion in statistical tests for performance comparison. In this sense, we can compare the different algorithms pairwise, regardless of their price range in dollars. Moreover, different series present different difficulty levels for forecasting. Therefore, we will not dive into details in summarizing the performance metrics for all the $148$ series considered in this study. Nevertheless, we present a case study for the stock APH, which generated an odd behavior in the TU coefficient during our analysis.

\subsection{Statistical comparison between the algorithms}

Firstly, we discuss our obtained results concerning MSE. This analysis is presented in Figure~\ref{fig:nemenyi_mse}. In the figure, the algorithms are ranked according to their forecasting errors. The most accurate algorithms have the smallest ranks. Algorithms that do not statistically differ in their resulting MSE (with $\alpha = 0.05$) are connected with a horizontal bar. As clearly depicted in the image, the MegazordNet variants occupied the first positions. The first connected group was composed by the CNN-based variants, whereas the LSTM ones composed the second one. Interestingly, regardless of the algorithm for trend composition, the season component did not appear to influence in great extents the ranking of the MegazordNet variants. In all of the cases, the models that only used trend predictors not differed from their seasonal predictor equipped counterparts. Nonetheless, in such applications, every cent worth of accuracy matters. Therefore, we advise using $\text{MegazordNet}_{\text{C,C}}$ when small MSE is the first concern.

\input{figs/nemenyi/nemenyi_mse.tex}

Among the traditional algorithms for time series prediction, the autoregressive models and SES were placed in the third most accurate algorithm group. Among them, RW generated the smallest MSE, despite not presenting statistically relevant improvements against its competitors. Considering the random characteristic of the latter method and its ranking, it seems that neither of the statistical-based algorithms was able to satisfactorily capturing the movements in the evaluated stocks. Both ARIMA and SES reached almost the same ranking. The most straightforward AR was the least accurate algorithm among the autoregressive ones. Lastly, both k-NN-TSP and MA were considered statistically equivalent accordingly with the performed test.

The TU coefficient compares each algorithm against a trivial baseline predictor, being this baseline the observation of the previous day. In this sense, this metric is very effective when coupled with MSE, for instance, to measure the extent to which an algorithm was able to capture the movements of the time series. The smaller the TU, the higher the performance gain obtained by the considered algorithm. We present the statistical test results for TU in Figure~\ref{fig:nemenyi_tu}. Again, the same ranking was observed among the MegazordNet variants. The CNN-based models achieved the best values of TU, while the LSTM ones again reached the second best positions. All the MegazordNet variants using the same type of neural network for the trend component were grouped. 

However, the rank positions changed for the traditional TS forecasting algorithms. RW, which appeared as the best competitor among the statistical solutions considering MSE, was ranked in the last position in this analysis. This fact was expected, since this solution basically applies a random deviation from the past state, i.e., the last observed time step. In this sense, on average, it performs worse than just replicating the last observation. In general, the autoregressive models tended to replicate the last day observation plus some degree of deviation. In this analysis, SES was the best traditional approach, followed by k-NN-TSP and MA.

\input{figs/nemenyi/nemenyi_tu.tex}

As a matter of illustration, we compared the mean TU value obtained by the MegazordNet variants for each TS, against the minimum TU value obtained for the other solutions. Therefore, our solutions were aggregated and compared against the best competitor for each TS considered. This analysis is presented in Figure~\ref{fig:tu_comparison}. As it can be seen in the chart, even without taking our best model, MegazordNet was able to surpass the best among the traditional forecasting algorithms in the great majority of cases. In fact, excluding the stock APH, MegazordNet in the worst case tied with its best competitor (in the stock CTL). We will delve into details into the specific case where our proposal performed worse than its competitors later. Interestingly, the traditional TSF algorithms in the majority of the cases performed very similarly to the naive predictor, since their TU was near to $1$ (refer to Section~\ref{sec:metrics} for details).

When considering POCID, the Megazord variants again reached the best positions in ranking, as showed in Figure~\ref{fig:nemenyi_pocid}. In this sense, our proposal was the best solution in predicting up and down trends in the stocks considered in this study. We observed a slight change in the ordering of our proposal's variants. However, given the small differences in the ranking and the nonexistence of statistically relevant difference among them, we cannot conclude that one variant is certainly better than the other regarding POCID.

Concerning the traditional TSF algorithms, the ranking positions also changed. MA reached the best ranking among the traditional algorithms. This situation was expected given that this technique ends to mostly modeling a smoothed version of the analyzed TS, as it used a window size of $20$ days. In fact, MA is the mechanism employed by the current version of MegazordNet to extract the trend component. SES and the autoregressive models appeared next in the ranking, followed by k-NN-TSP. Our experimental findings have shown that the autoregressive models tend to mimic the last observed day in their forecasting, which seemed to be the case with SES as well. The results of this forecasting behavior are to not satisfactorily capture the trends in the analyzed TS.

\input{figs/nemenyi/nemenyi_pocid.tex}

We also performed a comparison of our proposal's mean POCID performance against its best competitor, as depicted in Figure~\ref{fig:pocid_comparison}. As the chart shows, MegazordNet was the best performer when considering POCID, regardless of the considered TS. The mean POCID achieved by MegazordNet surpassed the $50\%$ mark in the majority of the cases. Therefore, our results are situated above the random guess strategy and superior, on average, to some other works~\citep{wen2019stock}. In the future, we intend to employ external sources of information, such as news and sentiment analysis about the economy, to enhance the capability of MegazordNet in predicting up and down trends in the stocks.

\begin{landscape}
\setlength{\abovecaptionskip}{5pt plus 3pt minus 2pt} % Chosen fairly arbitrarily
\begin{figure}[!htb]
    \centering
    \includegraphics[width=1.6\textwidth]{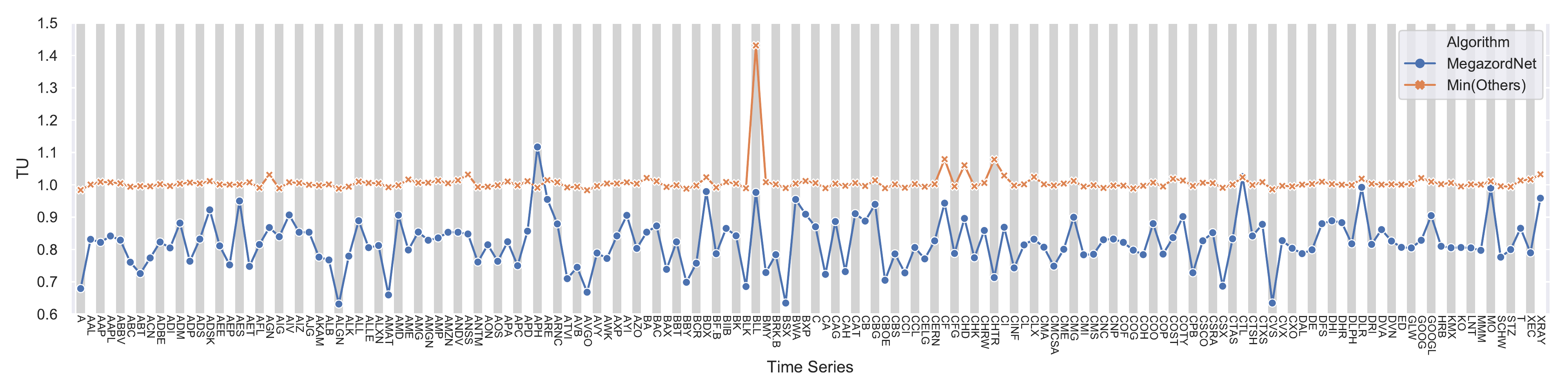}
    \caption{Comparing TU: mean performance of MegazordNet variants against the best traditional forecasting technique competitor}
    \label{fig:tu_comparison}
\end{figure}
\begin{figure}[!htb]
    \centering
    \includegraphics[width=1.6\textwidth]{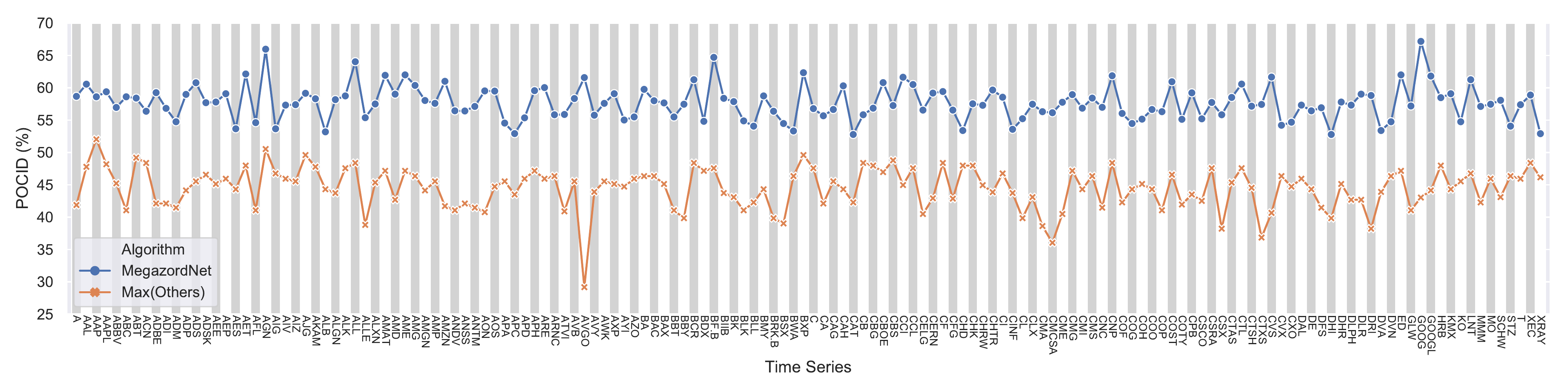}
    \caption{Comparing POCID: mean performance of MegazordNet variants against the best traditional forecasting technique competitor}
    \label{fig:pocid_comparison}
\end{figure}
\end{landscape}

\subsection{Case study: the APH stock}

\begin{figure}[!htb]
    \centering
    \subfloat[APH stock]{\includegraphics[width=\textwidth]{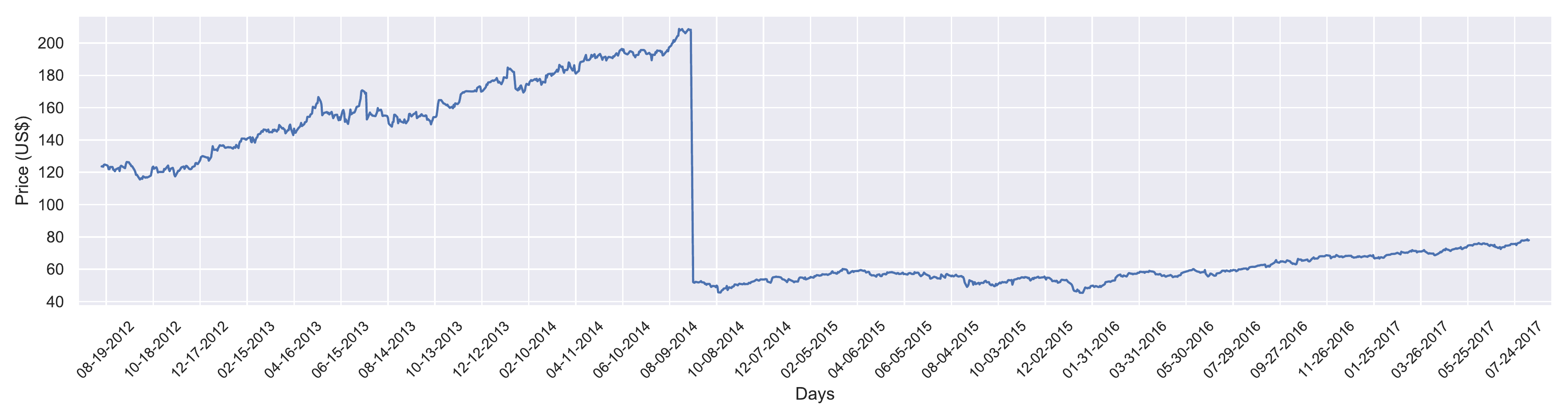}\label{fig:aph_all}}
    
    \subfloat[Detailing the sudden decrease]{\includegraphics[width=0.5\textwidth]{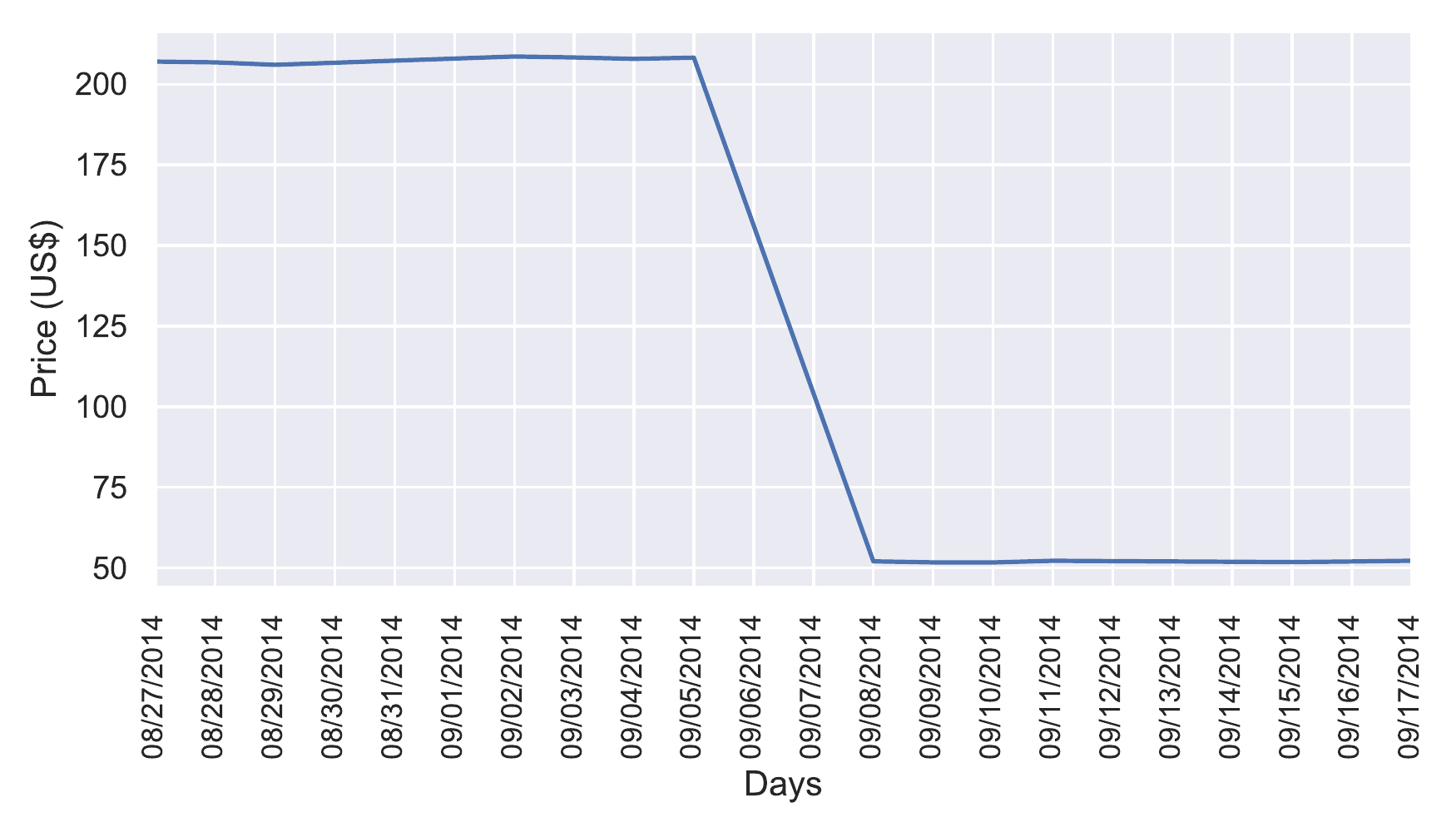}\label{fig:aph_zoom}}
    \subfloat[First derivative]{\includegraphics[width=0.5\textwidth]{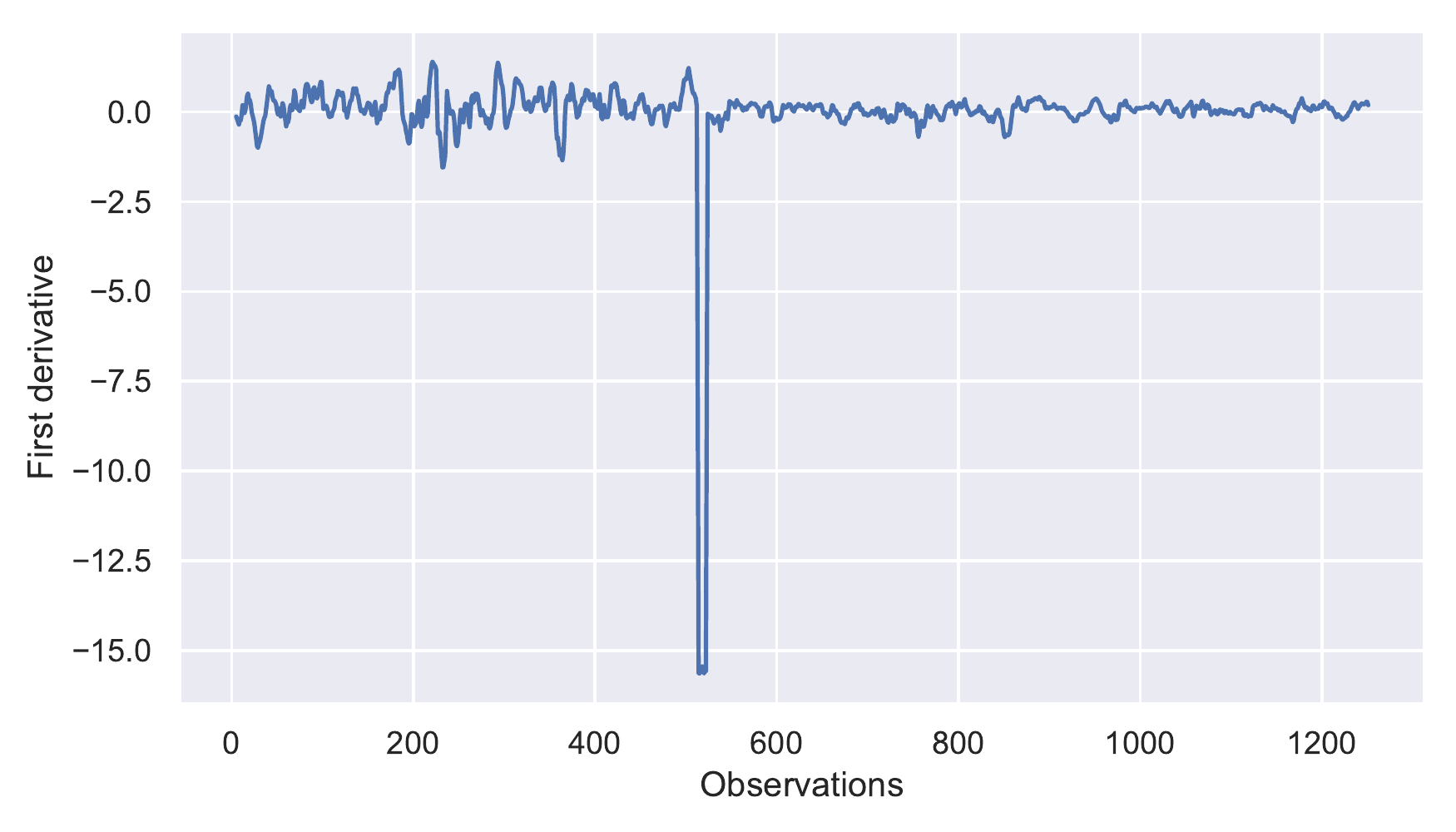}\label{fig:aph_1st_derivative}}
    
    \caption{APH stock: case where MegazordNets were surpassed by the traditional forecasting algorithms}
    \label{fig:aph_stock}
\end{figure}

Lastly, given that MegazordNet just lost in the stock of APH accordingly to its TU coefficient, we analyzed this specific company in details. The characteristics of the mentioned TS are presented in Figure~\ref{fig:aph_all}. As can be seen, there is a sudden decrease in the price of the stock around September of 2014. A zoomed view of this phenomena is presented in Figure~\ref{fig:aph_zoom}. In four days (from 09-05-2014 to 09-08-2014) the price of this stock decreased more than $100$ US\$, and any learning algorithm can hardly model this kind of situation. The resulting first derivative component, showed in Figure~\ref{fig:aph_1st_derivative}, illustrates this fact. This representation is employed by MegazordNet to learn the variations between the unitary intervals in the TS.

Seeking for the phenomena occurred in APH in other platforms, such as Yahoo Finance, we found that the observed decrease appears to be an inconsistency in the employed dataset. Therefore, more robust data extraction procedures must be employed when considering a real-world application of our proposal. Besides, considering that we did not employ an online learning mechanism in our experiments, MegazordNet was biased towards erroneous behaviors. This explains why it performed worse than its competitors in this specific TS. The remaining forecasting algorithms, on the other hand, employed online learning mechanisms, e.g., models based on rolling windows, hence being able to better adapt to the observed changes. Nevertheless, in all other evaluated cases, MegazordNet achieved the best results. The observed non-stationary characteristics in this specific TS motivated us to consider exploring online learning procedures for MegazordNet in the future.

%% file: figs/nemenyi/nemenyi_mse.tex
\begin{figure}[!htb]
  \centering
  \begin{tikzpicture}[xscale=2]
  \node (Label) at (0.8424347836903854, 0.7){\tiny{CD = 1.37}}; % the label
  \draw[decorate,decoration={snake,amplitude=.4mm,segment length=1.5mm,post length=0mm},very thick, color = black] (0.5,0.5) -- (1.1848695673807708,0.5);
  \foreach \x in {0.5, 1.1848695673807708} \draw[thick,color = black] (\x, 0.4) -- (\x, 0.6);

  \draw[gray, thick](0.5,0) -- (6.0,0);
  \foreach \x in {0.5,1.0,1.5,2.0,2.5,3.0,3.5,4.0,4.5,5.0,5.5,6.0} \draw (\x cm,1.5pt) -- (\x cm, -1.5pt);
  \node (Label) at (0.5,0.2){\tiny{1}};
  \node (Label) at (1.0,0.2){\tiny{2}};
  \node (Label) at (1.5,0.2){\tiny{3}};
  \node (Label) at (2.0,0.2){\tiny{4}};
  \node (Label) at (2.5,0.2){\tiny{5}};
  \node (Label) at (3.0,0.2){\tiny{6}};
  \node (Label) at (3.5,0.2){\tiny{7}};
  \node (Label) at (4.0,0.2){\tiny{8}};
  \node (Label) at (4.5,0.2){\tiny{9}};
  \node (Label) at (5.0,0.2){\tiny{10}};
  \node (Label) at (5.5,0.2){\tiny{11}};
  \node (Label) at (6.0,0.2){\tiny{12}};
  \draw[decorate,decoration={snake,amplitude=.4mm,segment length=1.5mm,post length=0mm},very thick, color = black](1.1155405405405405,-0.25) -- (1.4283783783783783,-0.25);
  \draw[decorate,decoration={snake,amplitude=.4mm,segment length=1.5mm,post length=0mm},very thick, color = black](2.1932432432432436,-0.25) -- (2.441891891891892,-0.25);
  \draw[decorate,decoration={snake,amplitude=.4mm,segment length=1.5mm,post length=0mm},very thick, color = black](3.95,-0.25) -- (4.472297297297297,-0.25);
  \draw[decorate,decoration={snake,amplitude=.4mm,segment length=1.5mm,post length=0mm},very thick, color = black](5.45,-0.25) -- (6.05,-0.25);
  \node (Point) at (1.1655405405405406, 0){};\node (Label) at (0.5,-1.05){\scriptsize{$\text{MegazordNet}_{\text{C,C}}$}}; \draw (Point) |- (Label);
  \node (Point) at (1.1858108108108107, 0){};\node (Label) at (0.5,-1.35){\scriptsize{$\text{MegazordNet}_{\text{C,L}}$}}; \draw (Point) |- (Label);
  \node (Point) at (1.3783783783783783, 0){};\node (Label) at (0.5,-1.65){\scriptsize{$\text{MegazordNet}_{\text{C,0}}$}}; \draw (Point) |- (Label);
  \node (Point) at (2.2432432432432434, 0){};\node (Label) at (0.5,-1.95){\scriptsize{$\text{MegazordNet}_{\text{L,L}}$}}; \draw (Point) |- (Label);
  \node (Point) at (2.2871621621621623, 0){};\node (Label) at (0.5,-2.25){\scriptsize{$\text{MegazordNet}_{\text{L,C}}$}}; \draw (Point) |- (Label);
  \node (Point) at (2.391891891891892, 0){};\node (Label) at (0.5,-2.55){\scriptsize{$\text{MegazordNet}_{\text{L,0}}$}}; \draw (Point) |- (Label);
  \node (Point) at (6.0, 0){};\node (Label) at (6.5,-1.05){\scriptsize{MA}}; \draw (Point) |- (Label);
  \node (Point) at (5.5, 0){};\node (Label) at (6.5,-1.35){\scriptsize{k-NN-TSP}}; \draw (Point) |- (Label);
  \node (Point) at (4.422297297297297, 0){};\node (Label) at (6.5,-1.65){\scriptsize{AR}}; \draw (Point) |- (Label);
  \node (Point) at (4.219594594594595, 0){};\node (Label) at (6.5,-1.95){\scriptsize{ARIMA}}; \draw (Point) |- (Label);
  \node (Point) at (4.206081081081081, 0){};\node (Label) at (6.5,-2.25){\scriptsize{SES}}; \draw (Point) |- (Label);
  \node (Point) at (4.0, 0){};\node (Label) at (6.5,-2.55){\scriptsize{RW}}; \draw (Point) |- (Label);
  \end{tikzpicture}
  \caption{Nemenyi test results considering MSE}
  \label{fig:nemenyi_mse}
\end{figure}

%% file: figs/nemenyi/nemenyi_tu.tex
\begin{figure}[!htb]
  \centering
  \begin{tikzpicture}[xscale=2]
  \node (Label) at (0.8424347836903854, 0.7){\tiny{CD = 1.37}}; % the label
  \draw[decorate,decoration={snake,amplitude=.4mm,segment length=1.5mm,post length=0mm},very thick, color = black] (0.5,0.5) -- (1.1848695673807708,0.5);
  \foreach \x in {0.5, 1.1848695673807708} \draw[thick,color = black] (\x, 0.4) -- (\x, 0.6);

  \draw[gray, thick](0.5,0) -- (6.0,0);
  \foreach \x in {0.5,1.0,1.5,2.0,2.5,3.0,3.5,4.0,4.5,5.0,5.5,6.0} \draw (\x cm,1.5pt) -- (\x cm, -1.5pt);
  \node (Label) at (0.5,0.2){\tiny{1}};
  \node (Label) at (1.0,0.2){\tiny{2}};
  \node (Label) at (1.5,0.2){\tiny{3}};
  \node (Label) at (2.0,0.2){\tiny{4}};
  \node (Label) at (2.5,0.2){\tiny{5}};
  \node (Label) at (3.0,0.2){\tiny{6}};
  \node (Label) at (3.5,0.2){\tiny{7}};
  \node (Label) at (4.0,0.2){\tiny{8}};
  \node (Label) at (4.5,0.2){\tiny{9}};
  \node (Label) at (5.0,0.2){\tiny{10}};
  \node (Label) at (5.5,0.2){\tiny{11}};
  \node (Label) at (6.0,0.2){\tiny{12}};
  \draw[decorate,decoration={snake,amplitude=.4mm,segment length=1.5mm,post length=0mm},very thick, color = black](1.0952702702702701,-0.25) -- (1.4182432432432432,-0.25);
  \draw[decorate,decoration={snake,amplitude=.4mm,segment length=1.5mm,post length=0mm},very thick, color = black](2.162837837837838,-0.25) -- (2.4182432432432432,-0.25);
  \draw[decorate,decoration={snake,amplitude=.4mm,segment length=1.5mm,post length=0mm},very thick, color = black](3.4263513513513515,-0.25) -- (4.05,-0.25);
  \draw[decorate,decoration={snake,amplitude=.4mm,segment length=1.5mm,post length=0mm},very thick, color = black](3.95,-0.4) -- (4.55,-0.4);
  \draw[decorate,decoration={snake,amplitude=.4mm,segment length=1.5mm,post length=0mm},very thick, color = black](4.45,-0.4) -- (5.212162162162162,-0.4);
  \draw[decorate,decoration={snake,amplitude=.4mm,segment length=1.5mm,post length=0mm},very thick, color = black](5.1121621621621625,-0.25) -- (5.769594594594594,-0.25);
  \node (Point) at (1.1452702702702702, 0){};\node (Label) at (0.5,-1.4500000000000002){\scriptsize{$\text{MegazordNet}_{\text{C,C}}$}}; \draw (Point) |- (Label);
  \node (Point) at (1.1655405405405406, 0){};\node (Label) at (0.5,-1.7500000000000002){\scriptsize{$\text{MegazordNet}_{\text{C,L}}$}}; \draw (Point) |- (Label);
  \node (Point) at (1.3682432432432432, 0){};\node (Label) at (0.5,-2.0500000000000003){\scriptsize{$\text{MegazordNet}_{\text{C,0}}$}}; \draw (Point) |- (Label);
  \node (Point) at (2.2128378378378377, 0){};\node (Label) at (0.5,-2.35){\scriptsize{$\text{MegazordNet}_{\text{L,L}}$}}; \draw (Point) |- (Label);
  \node (Point) at (2.2635135135135136, 0){};\node (Label) at (0.5,-2.6500000000000004){\scriptsize{$\text{MegazordNet}_{\text{L,C}}$}}; \draw (Point) |- (Label);
  \node (Point) at (2.3682432432432434, 0){};\node (Label) at (0.5,-2.95){\scriptsize{$\text{MegazordNet}_{\text{L,0}}$}}; \draw (Point) |- (Label);
  \node (Point) at (5.719594594594594, 0){};\node (Label) at (6.5,-1.4500000000000002){\scriptsize{RW}}; \draw (Point) |- (Label);
  \node (Point) at (5.618243243243243, 0){};\node (Label) at (6.5,-1.7500000000000002){\scriptsize{ARIMA}}; \draw (Point) |- (Label);
  \node (Point) at (5.162162162162162, 0){};\node (Label) at (6.5,-2.0500000000000003){\scriptsize{AR}}; \draw (Point) |- (Label);
  \node (Point) at (4.5, 0){};\node (Label) at (6.5,-2.35){\scriptsize{MA}}; \draw (Point) |- (Label);
  \node (Point) at (4.0, 0){};\node (Label) at (6.5,-2.6500000000000004){\scriptsize{k-NN-TSP}}; \draw (Point) |- (Label);
  \node (Point) at (3.4763513513513513, 0){};\node (Label) at (6.5,-2.95){\scriptsize{SES}}; \draw (Point) |- (Label);
  \end{tikzpicture}
  \caption{Nemenyi test results considering TU}
  \label{fig:nemenyi_tu}
\end{figure}

%% file: figs/nemenyi/nemenyi_pocid.tex
\begin{figure}[!htb]
  \centering
  \begin{tikzpicture}[xscale=2]
  \node (Label) at (0.8424347836903854, 0.7){\tiny{CD = 1.37}}; % the label
  \draw[decorate,decoration={snake,amplitude=.4mm,segment length=1.5mm,post length=0mm},very thick, color = black] (0.5,0.5) -- (1.1848695673807708,0.5);
  \foreach \x in {0.5, 1.1848695673807708} \draw[thick,color = black] (\x, 0.4) -- (\x, 0.6);

  \draw[gray, thick](0.5,0) -- (6.0,0);
  \foreach \x in {0.5,1.0,1.5,2.0,2.5,3.0,3.5,4.0,4.5,5.0,5.5,6.0} \draw (\x cm,1.5pt) -- (\x cm, -1.5pt);
  \node (Label) at (0.5,0.2){\tiny{1}};
  \node (Label) at (1.0,0.2){\tiny{2}};
  \node (Label) at (1.5,0.2){\tiny{3}};
  \node (Label) at (2.0,0.2){\tiny{4}};
  \node (Label) at (2.5,0.2){\tiny{5}};
  \node (Label) at (3.0,0.2){\tiny{6}};
  \node (Label) at (3.5,0.2){\tiny{7}};
  \node (Label) at (4.0,0.2){\tiny{8}};
  \node (Label) at (4.5,0.2){\tiny{9}};
  \node (Label) at (5.0,0.2){\tiny{10}};
  \node (Label) at (5.5,0.2){\tiny{11}};
  \node (Label) at (6.0,0.2){\tiny{12}};
  \draw[decorate,decoration={snake,amplitude=.4mm,segment length=1.5mm,post length=0mm},very thick, color = black](1.3097972972972973,-0.25) -- (1.4317567567567568,-0.25);
  \draw[decorate,decoration={snake,amplitude=.4mm,segment length=1.5mm,post length=0mm},very thick, color = black](2.0614864864864866,-0.25) -- (2.25777027027027,-0.25);
  \draw[decorate,decoration={snake,amplitude=.4mm,segment length=1.5mm,post length=0mm},very thick, color = black](4.161148648648648,-0.25) -- (4.84222972972973,-0.25);
  \draw[decorate,decoration={snake,amplitude=.4mm,segment length=1.5mm,post length=0mm},very thick, color = black](4.705067567567568,-0.4) -- (5.129391891891892,-0.4);
  \node (Point) at (1.3597972972972974, 0){};\node (Label) at (0.5,-1.05){\scriptsize{$\text{MegazordNet}_{\text{C,L}}$}}; \draw (Point) |- (Label);
  \node (Point) at (1.3631756756756757, 0){};\node (Label) at (0.5,-1.35){\scriptsize{$\text{MegazordNet}_{\text{C,0}}$}}; \draw (Point) |- (Label);
  \node (Point) at (1.3817567567567568, 0){};\node (Label) at (0.5,-1.65){\scriptsize{$\text{MegazordNet}_{\text{C,C}}$}}; \draw (Point) |- (Label);
  \node (Point) at (2.1114864864864864, 0){};\node (Label) at (0.5,-1.95){\scriptsize{$\text{MegazordNet}_{\text{L,0}}$}}; \draw (Point) |- (Label);
  \node (Point) at (2.1841216216216215, 0){};\node (Label) at (0.5,-2.25){\scriptsize{$\text{MegazordNet}_{\text{L,L}}$}}; \draw (Point) |- (Label);
  \node (Point) at (2.20777027027027, 0){};\node (Label) at (0.5,-2.55){\scriptsize{$\text{Megazord}_{\text{L,C}}$}}; \draw (Point) |- (Label);
  \node (Point) at (5.079391891891892, 0){};\node (Label) at (6.5,-1.05){\scriptsize{k-NN-TSP}}; \draw (Point) |- (Label);
  \node (Point) at (4.79222972972973, 0){};\node (Label) at (6.5,-1.35){\scriptsize{RW}}; \draw (Point) |- (Label);
  \node (Point) at (4.787162162162162, 0){};\node (Label) at (6.5,-1.65){\scriptsize{AR}}; \draw (Point) |- (Label);
  \node (Point) at (4.766891891891892, 0){};\node (Label) at (6.5,-1.95){\scriptsize{ARIMA}}; \draw (Point) |- (Label);
  \node (Point) at (4.7550675675675675, 0){};\node (Label) at (6.5,-2.25){\scriptsize{SES}}; \draw (Point) |- (Label);
  \node (Point) at (4.211148648648648, 0){};\node (Label) at (6.5,-2.55){\scriptsize{MA}}; \draw (Point) |- (Label);
  \end{tikzpicture}
  \caption{Nemenyi test results considering POCID}
  \label{fig:nemenyi_pocid}
\end{figure}

%% file: sections/06_final_considerations.tex
\section{Final Considerations}\label{final-considerations}

In this study, we presented a novel framework called MegazordNet for FTSF that combine statistical analysis and artificial neural networks. Being the current work a preliminary study, we selected at random $148$ TS from the S\&P 500 index for experimental evaluation. Despite being simple in its core design regarding the employed data transformation procedures, MegazordNet was able to statistically surpassing traditional statistical and ML-based algorithms for TSF, regardless of the performance metric considered. 

There are some key-point designs of our work that can be expanded. For instance, we may test some different model combinations using temporal causal convolutions or attention-based networks in order to explore the combination of other related exogenous time-series. 
Additionally, we did not use a multivariate TSF approach or any modeling for the residual component, which could easily give our model a more generalized perception of the stock market.
In the future, we also intend to evaluate the impact of white noise addition in our predictive performance, as well as, external sources of information (e.g., sentiment analysis) as inputs for the MegazordNet, since its architecture is generic and expandable. 